\documentclass[traditabstract]{aa} 
\usepackage{graphicx}
\usepackage{subfigure}
\usepackage{subfig}
\usepackage{natbib}
\usepackage{longtable}
\usepackage[english]{babel}
\usepackage{txfonts}
\def \xmmn {\emph{XMM-Newton }}
\def \xmm {\emph{XMM-Newton}}
\def \chandra {\emph{Chandra }}

\begin{document}

\title{Back and forth from cool core to non-cool core: clues from radio-halos}
\author{M.\,Rossetti
         \inst{1,2}
	\and 
	D.\,Eckert\inst{1}
                  \and
         B.\,M.\,Cavalleri\inst{2,1}
	\and
	S.\,Molendi\inst{1}
	\and F.\,Gastaldello\inst{1,3}
  \and S.\,Ghizzardi\inst{1}       }

\institute{Istituto di Astrofisica Spaziale e Fisica Cosmica, INAF, via Bassini 15, I-20133 Milano, Italy
\and
Dipartimento di Fisica -- Universit\`a degli Studi di Milano
via Celoria 16, Milano, Italy
\and
University of California at Irvine, 4129, Frederick Reines Hall, Irvine, CA
92697-4575, USA}

\abstract{X-ray astronomers often divide galaxy clusters 
 into two classes: ``cool core'' (CC) and ``non-cool core'' (NCC) objects.
The origin of this dichotomy has been the subject of debate in recent years, between
 ``evolutionary'' models (where clusters can evolve from CC to NCC, mainly through
 mergers) and ``primordial'' models (where the state of the cluster is fixed ``ab initio'' by early mergers or pre-heating).
We found that in a well-defined sample (clusters in the GMRT Radio halo survey with available \chandra or \xmmn data), none of the objects hosting a giant radio halo can be classified as a cool core.
This result suggests that the main mechanisms which can start a large scale synchrotron emission (most likely mergers) are the same that can destroy CC  and therefore strongly supports ``evolutionary'' models of the CC-NCC dichotomy. Moreover combining the number of objects in the CC and NCC state with the number of objects with and without a radio-halo, we estimated that the time scale over which a NCC cluster relaxes to the CC state, should be larger than the typical life-time of radio-halos and likely shorter than $\simeq 3$ Gyr. This suggests that NCC transform into CC more rapidly than predicted from the cooling time, which is about $10$ Gyr in NCC systems, allowing the possibility of a cyclical evolution between the CC and NCC states.
}
\maketitle

\section{Introduction}

Galaxy clusters are often divided by X-ray astronomers into two classes: ``cool core''(CC) and ``non-cool core'' (NCC) objects on the basis of the observational properties of their central regions. 
One of the open questions in the study of galaxy clusters concerns the origin of this distribution. 
The original model which prevailed for a long time assumed that the CC state was a sort of ``natural state'' for the clusters, and the observational features were explained within the old ``cooling flow'' model: radiation losses cause the gas in the centers of these clusters to cool and to flow inward. Clusters were supposed to live in this state until disturbed by a ``merger''. Indeed, mergers are very energetic events that can shock-heat \citep{burns97} and mix the ICM \citep{gomez02}: through these processes they were supposed to efficiently destroy cooling flows (e.\,g.\, \citealt{mcglynn84}). After the mergers, clusters were supposed to relax and go back to the cooling flow state in a sort of cyclical evolution (e.\,g.\, \citealt{buote02}).
With the fall of the ``cooling flow'' model brought about by the \xmmn and \chandra observations (e.\,g.\, \citealt{peterson01}), doubts were cast also on the interpretation of mergers as the dominant mechanism which could transform CC clusters into NCC. These doubts were also motivated by the difficulties of numerical simulations in destroying simulated cool cores with mergers (e.\,g.\, \citealt{burns08} and references therein).
More generally speaking, the question arose whether the observed distribution of clusters was due to a primordial division into the two classes or rather to evolutionary differences during the history of the clusters.  \\
For instance \citet{mccarthy04, mccarthy08} suggested that early episodes of non-gravitational pre-heating in the redshift range $1<z<2$ may have increased the entropy of the ICM of some proto-clusters which did not have time to develop a full cool core. \citet{burns08} suggested that while mergers cannot destroy simulated cool cores in the local Universe, early major mergers could have destroyed nascent cool cores in an earlier phase of their formation $(z>0.5)$. \\
However, the ``evolutionary'' scenario, where recent and on-going mergers are responsible for the CC-NCC dichotomy, has been continuously supported by observations. Indeed, correlations have been found between the lack of a cool core and several multi-wavelength indicators of on-going dynamical activity (e.\,g.\, \citealt{sanderson06,sanderson09b}, \citealt{leccardi10}). \\
Giant radio halos (RH) are the most spectacular evidence of non thermal emission in galaxy clusters (\citealt{ferrari08} for a recent review). Over the last years, there has been increasing evidence in the literature that they are found in clusters with a strong on-going dynamical activity (e.\,g.\,\citealt{buote01, schuecker01},  \citealt{hudson10}) suggesting that mergers could provide the energy necessary to accelerate (or re-accelerate) electrons to radio-emitting energies \citep{sarazin02,brunetti09}. Recently, the connection between radio halos and mergers has been quantitatively confirmed on a well-defined statistical sample by \citet{cassano10}. \\
In the framework of ``evolutionary'' scenarios, mergers are also responsible for the CC-NCC dichotomy. Therefore, we expect mergers to cause a relation between the absence of a cool core and the presence of a giant radio halo. The aim of this work is to assess statistically the presence of this relation on a well-defined sample and to test our ``evolutionary'' interpretation of the origin of the CC-NCC distribution.\\
Throughout this paper we assume a $\Lambda$CDM cosmology with $H_0=70\, \rm{km\,s^{-1}\,Mpc^{-1}}$, $\Omega_M=0.3$ and $\Omega_\Lambda=0.7$.


\section{The sample}
\label{sec_sample}
The choice of the sample is an important part of this project. We do not want to select an ``archive radio sample'' of clusters, since the search for radio halos often concentrated in clusters which showed some indications of a disturbed dynamical state from other wavelengths, such as the absence of a cool core in X-rays. Unlike archival samples, the ``GMRT radio halo survey'' \citep{venturi07, venturi08} is an excellent starting point for our aims: it consists of a deep pointed radio survey of a representative sample of 50 clusters selected in X-rays from flux-limited  \emph{ROSAT} surveys (REFLEX and eBCS), with $z=0.2-0.4$, $L_X>5\times10^{44}\, \rm{ergs}\,\rm{s}^{-1}$ and $-30\degr<\delta<60\degr$. For 35 clusters of this sample, uniformly observed with GMRT, \citet{venturi08} could either detect extended radio emission or put strong upper limits on it. 
The GMRT RH sample is designed to cover a well defined range in X-ray luminosity, such that according to the $P_{1.4\,GHz}-L_X$ relation (e.\,g.\, \citealt{brunetti09}), any radio-halo should have been detected in the survey. Therefore we can consider ``radio-quiet'' the clusters for which \citet{venturi08} did not find extended central radio emission.\\
We looked into the \chandra and \xmmn archives for observations of the clusters in the GMRT RH sample. We preferentially used \chandra observations to exploit the better angular resolution but we discarded observations with less than 1500 counts in each of the regions from which we extract spectra (see Sec.\,\ref{CC-est}), moving to \xmmn when available. \\
To avoid confusion, we do not consider here the three objects in the GMRT RH survey with mini radio halos (A\,2390, RXCJ\,$1504.1-0248$ and Z\,7160). Their central radio emission has been classified as a mini-halo because it extends for less than 500 kpc and it is associated to an active central radio galaxy \citep{bacchi03,mazzotta08,giacintucci11}. Mini-halos should not be confused with giant RHs since their origin is likely due to electrons injected by the central AGN re-accelerated by  sloshing \citep{mazzotta08,zuhone11} or AGN-driven turbulence \citep{cassano08}.\\
Our final sample consists of 22 clusters with available \chandra or \xmmn observations (Table \ref{table}). We verified this subsample to be representative of the starting sample (50 objects) in terms of luminosity with a Kolmogorov--Smirnov test.
10 clusters in our subsample are ``radio-loud'' (hosting a giant radio halo obeying the well known relation between the radio power at $1.4$ GHz, $P_{1.4}$, and the X--ray luminosity $L_X$) and the remaining 12 are ``radio-quiet'',  well separated in the $P_{1.4}-L_X$ plane (see \citealt{brunetti09} for a detailed discussion of this distribution). We note here that our sample of ``radio--loud'' clusters is composed of all the clusters with a confirmed giant radio halo in \citet{venturi08}, with the addition of A697 and A1758 which were classified as ``candidate halos'' and were confirmed later \citep{macario10,giovannini09}.\\
We verified that in our sample clusters with radio-halos are neither more luminous in X-rays nor more distant than radio-quiet clusters.
\begin{table}[]
\caption{Cool core indicators for the clusters of the sample.}              
\label{table}      
\centering                                      
\begin{tabular}{c c c c}          
\hline\hline          

Number & Cluster & $K_0(\rm{keV\,cm}^2)$ & $\sigma$ \\
\hline
1 & A\,2163 &	$437.98 \pm	82.56 $	& $0.63 \pm 	0.03$	\\		
2 &A\,521\tablefootmark{*}	& $259.87 \pm	36.25 $ &	$0.57 \pm	0.05$ \\		
3 &A\,2219 & $411.57 \pm	43.16$ &	$0.77 \pm 0.06$	\\		
4 &A\,2744 & $438.44 \pm	58.71$ & $0.90 \pm	0.14 $	\\		
5 &A\,1758 & $230.84 \pm	37.22$ & $0.46 \pm 	0.04 $	\\
6 &RXCJ\,$2003.5-2323$\tablefootmark{\dag}	& $340.61 \pm 28.62$ & $0.91 \pm	0.10$	\\	
7 &A\,1300\tablefootmark{*}& $97.26 \pm	22.98$ & $0.54 \pm	0.06$ \\		
8 &A\,773 & $244.32 \pm	31.73$ & $0.71 \pm 	0.08$	\\		
9 &A\,697\tablefootmark{\dag}& $178.09 \pm	28.62$ & $0.71 \pm 	0.10$	\\		
10 &A\,209\tablefootmark{*} & $105.50 \pm	26.94$ & $0.58 \pm	0.04$ \\		
\hline										
11 &A\,1423 & $68.32 \pm 12.85 $ & $0.41 \pm 0.05$ \\			
12 & A\,2537 & $110.41 \pm 19.37 $ &	$0.36 \pm 0.04$ \\	
13 & A\,2631\tablefootmark{*}& $308.81 \pm	37.38 $ & $0.8 \pm	0.1$ \\		
14 &A\,2667 & $19.32 \pm	3.39$ & $0.32 \pm 	0.02 $ \\		
15 &A\,3088	& $82.78 \pm	8.42$ & $0.30 \pm 	0.03 $ \\	
16 &A\,611	& $124.93 \pm	18.61$& $0.47 \pm 	0.03$	\\		
17 &RXCJ\,$1115.8+0129$ & $14.76 \pm	3.09$	&	$0.26 \pm 0.02$ \\	
18 &RXJ\,$1532.9+3021$ & $16.93 \pm	1.81$ & $0.34 \pm 0.03$	\\		
19 &RXJ\,$2228.6+2037$\tablefootmark{*} & $118.79 \pm	39.19$ & $0.61 \pm 0.06$\\		
20 &Z\,2701& $39.66 \pm	3.92$	& $0.36 \pm 	0.01$	\\		
21 &Z\,2089\tablefootmark{\dag}& $14.59 \pm 1.14$	& $0.36 \pm	0.02$ \\		
22 &AS\,0780\tablefootmark{\dag} & $22.24 \pm 1.36$	& $0.26 \pm 0.01$ \\
\hline
\end{tabular}
\tablefoot{
The first column refers to the numbering of the objects in Fig.\,\ref{fig_ind}.
Clusters 1-10 are ``radio-halos'' while objects 11-22 are ``radio-quiet''.\\
\tablefoottext{*} {Pseudo-entropy ratio measured with \xmmn data.}
\tablefoottext{\dag} {\chandra observation not in the ACCEPT archive.}
}
\end{table}
\section{Data analysis}
\subsection{Chandra data reduction}
We analyzed \chandra data with the software CIAO v.$4.1$ and the calibration database (CALDB) $4.1.1$. All data were reprocessed from the "level 1" event files, following the standard \chandra data-reduction threads\footnote{http://cxc.harvard.edu/ciao/threads/index.html}.
We applied the standard corrections to  account for a time-dependent drift in the detector gain and charge transfer inefficiency, as implemented in the CIAO tools. From low surface brightness regions of the active chips we extracted a light-curve ($5-10$ keV) to identify and excise periods of enhanced background. 
We removed point sources detected with the CIAO tool {\verb wavdetect }.
Background analysis was performed using the blank-sky datasets provided in the CALDB. Background files were reprocessed and reprojected to match each observation. We extracted spectra from the background and source files from an external region not contaminated by cluster emission and we quantified the ratio between the count rate of the observation and of the background in a hard energy band ($9.5-12$ keV). By rescaling the background files for this number, we took into account 
possible temporal variations of the instrumental background dominating at high energies. This procedure does not introduce substantial distortions in the soft energy band, where the cosmic background components are more important, since we limit our analysis to regions where the source outshines the background in the soft band by more than one order of magnitude.

\subsection{XMM-Newton data reduction}
We have analyzed \xmmn observations for the 6 clusters (see Table \ref{table}) of our sample where we found a total number of \chandra counts $<1500$ in the $IN$ region (Sec.\,\ref{sigma_sec}) used to measure the pseudo-entropy ratios \citep{leccardi10}. We generated calibrated event files using the SAS software v.\,$9.0$ and then we removed soft proton flares using a double filtering process, first in a hard ($10-12$ keV) and then in a soft ($2-5$ keV) energy range. The event files were filtered according to {\verb PATTERN } and {\verb FLAG } criteria. Bright point-like sources were detected using a procedure based on the SAS task  {\verb edetect_chain } and removed from the event files. As background files, we merged nine blank field observations,  we reprojected them to match each observation and renormalized by the ratio of the count-rates in an external region as for the \chandra case (more details in \citealt{leccardi10}).
We extracted spectra for the three EPIC detectors  and fitted them separately. After verifying that best fit parameters from each instrument are consistent at less than $2\sigma$, we combined them with a weighted mean.

\subsection{Cool core indicators}
\label{CC-est}
For each of the clusters in our sample we have calculated two estimators of the core state: the central entropy $K_0$ (\citealt{cava09}, Sec.\,\ref{sec_k0}) and the pseudo-entropy ratio $\sigma$ (\citealt{leccardi10}, Sec.\,\ref{sigma_sec}).
To measure them, both with \chandra and \xmm,  we extracted spectra and generated an effective area (ARF) for each region, which we associated to the appropriate response file (RMF).
We fitted spectra within XSPEC $v.12$ with an absorbed {\it mekal} model, where we fixed redshift as given by NED\footnote{http://nedwww.ipac.caltech.edu/} and $N_H$ to the \citet{dlmap} values (for consistency with \citealt{cava09}). We verified that the $N_H$ values agree with those derived from the LAB map \citep{kalberla05} within 20\% for all but one cluster of the sample. For this cluster, temperatures and normalizations obtained with the two different values of $N_H$ are consistent at less than $1\sigma$. 

\subsubsection{Central Entropy $K_0$}
\label{sec_k0}
$K_0$ is derived from the fit of the entropy profile with the model $K_0+K_{100}(r/100\,\rm{kpc})^{\alpha}$.
When available, we have used the values reported in the ACCEPT catalogue\footnote{http://www.pa.msu.edu/astro/MC2/accept/} \citep{cava09}. For the 4 objects (see Table \ref{table}) whose \chandra observations were not public at the time of the compilation of ACCEPT, we have extracted the entropy profile following the same procedure as \citet{cava09}. More specifically, we combined the temperature profile, measured by fitting a thermal model to spectra extracted in concentric annuli with at least 2500 counts, with the gas density profile. The latter was derived using the deprojection technique of \citet{kriss83} by combining the surface brightness profile with the spectroscopic count rate and normalization in each region of the spectral analysis. Errors were estimated with a Monte Carlo simulation. \\ 
\begin{figure*}[!ht]
\subfigure{
\resizebox{0.5\hsize}{!}{\includegraphics[clip=true]{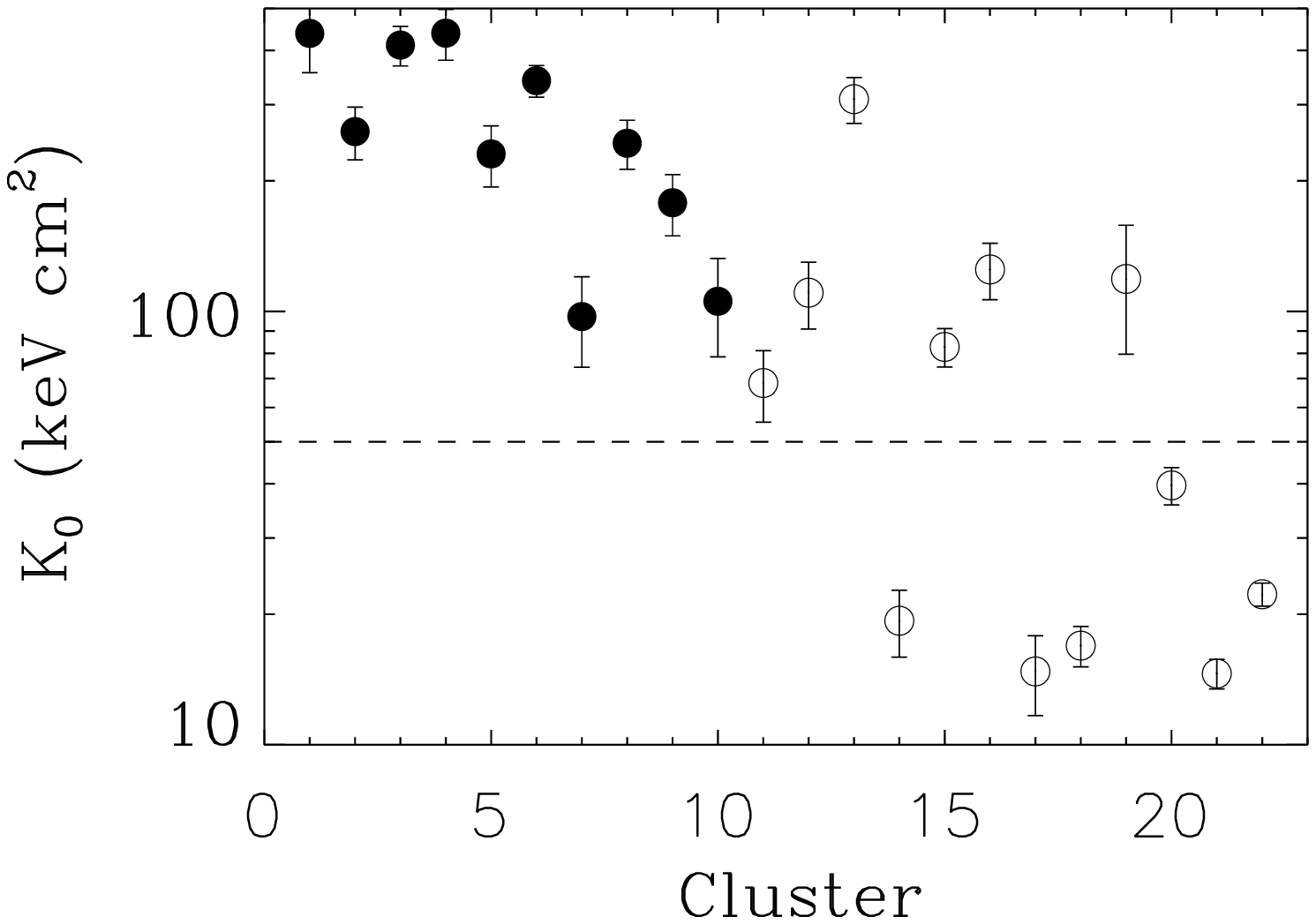}}
}
\subfigure{
\resizebox{0.5\hsize}{!}{\includegraphics[clip=true]{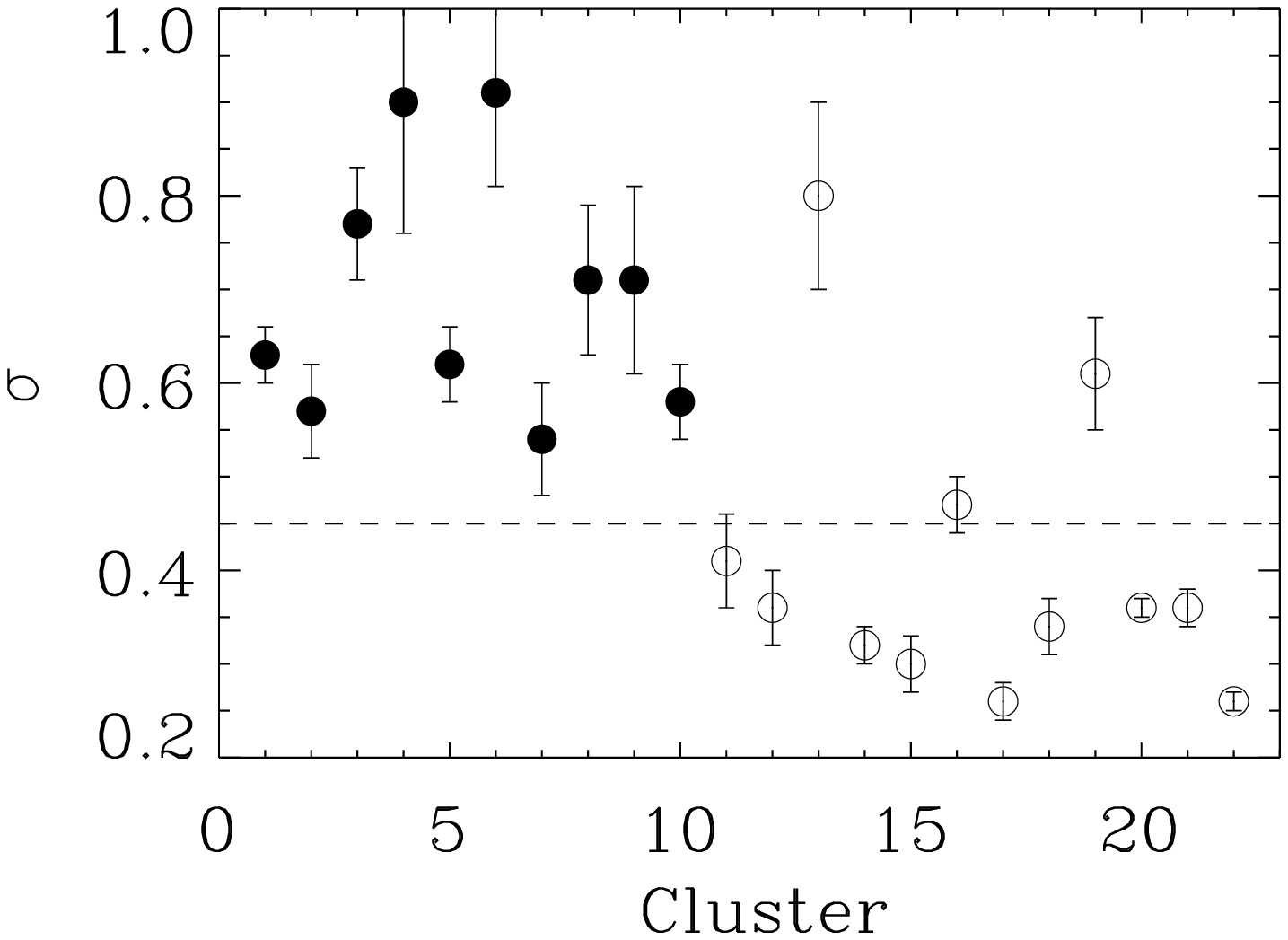}}
}
\caption{Cool core indicators ($K_0$ and $\sigma$) for all the clusters in the sample. Filled symbols are ``radio-halo'' clusters while open symbols are ``radio-quiet''. Dashed lines mark the threshold between CC (below) and NCC (above). The numbering on the ``x'' axis refers to the position of the objects in Table \ref{table}. Error bars are $1 \sigma$.}
\label{fig_ind}
\end{figure*}
\subsubsection{Pseudo-entropy ratios $\sigma$}
\label{sigma_sec}
The pseudo-entropy ratio is defined as $\sigma=(T_{IN}/T_{OUT})*(EM_{IN}/EM_{OUT})^{-1/3}$, where $T$ is the temperature, $EM$ is the emission measure (XSPEC normalization of the {\it mekal} model divided by the area of the region). The $IN$ and $OUT$ regions are defined with fixed fractions of $R_{180}$ ($R<0.05R_{180}$ for the $IN$ region and $0.05R_{180}<R<0.2R_{180}$ for the $OUT$ region). We calculated $R_{180}$ from $T_{OUT}$ using the expression in \citet{leccardi08a}, iterating the process until it converged to stable values of $R_{180}$ \citep{rossetti10}. The center from which we defined our region is the same used in ACCEPT (i.\,e.\, the X--ray peak or the centroid of the X-ray emission if these two points differ for more than 70 kpc).\\
The limited spatial resolution of EPIC may be an issue for measuring $\sigma$ in clusters at $z>0.25$, since it may cause the spreading of photons coming from the $IN$ circle into the $OUT$ region (and vice-versa, although this contamination is likely to be less important especially in CC). Therefore, we adopted the cross-talk modification of the ARF generation software and fitted simultaneously the spectra of the two regions (see \citealt{snowden08} and \citealt{ettori10} for details).


\section{Results}
\label{sec_results}
\begin{figure}[!t]
\hspace{-2.0 cm}
\subfigure{\includegraphics[width=6cm]{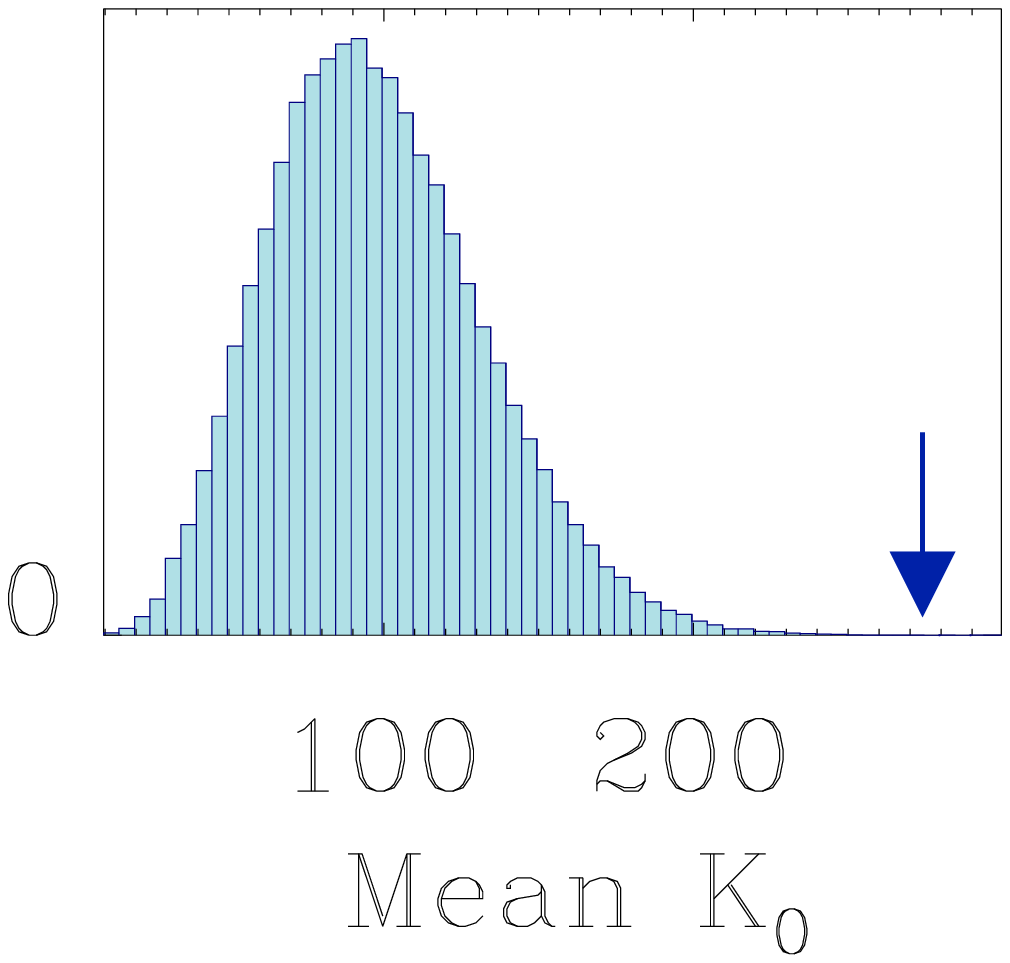}}
\hspace{-2.0 cm}
\subfigure{\includegraphics[width=6cm]{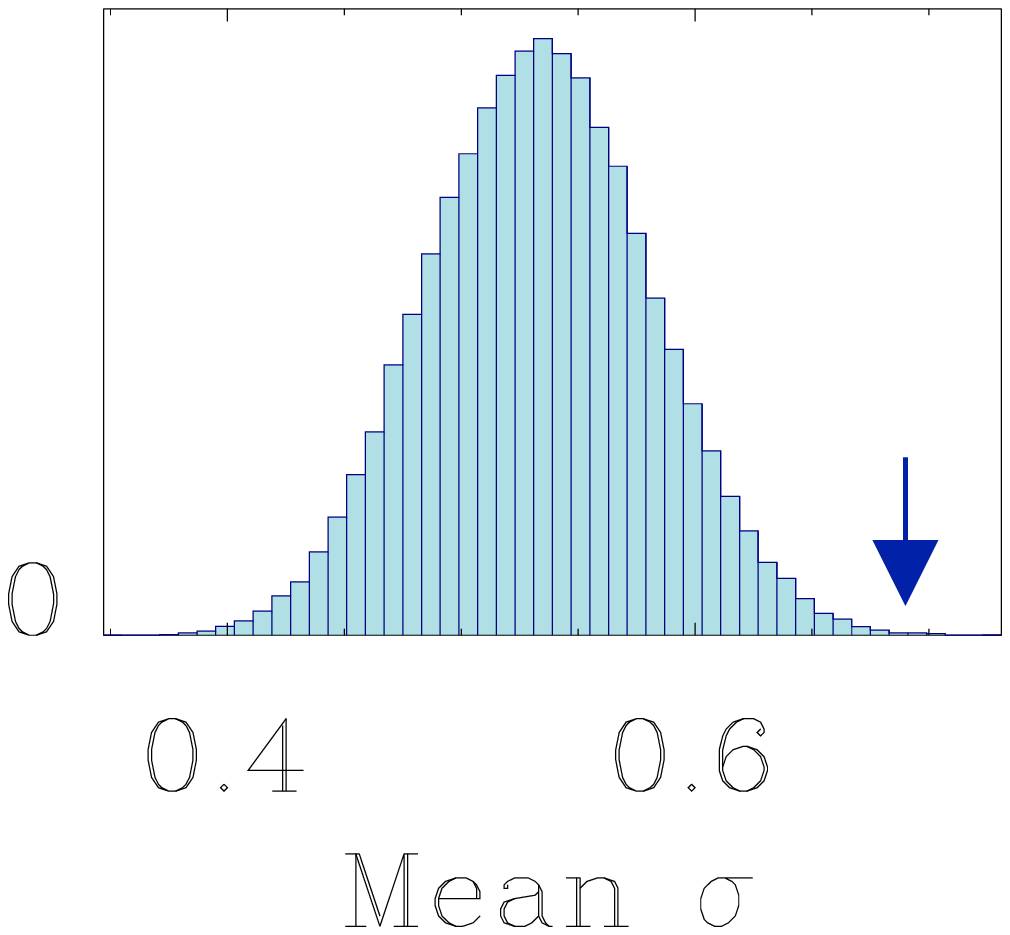}}
\caption{Histogram density for the mean $K_0$ (left) and $\sigma$ (right) of 10 randomly selected clusters in $10^5$ Monte Carlo simulations. The arrows show the observed mean values in our radio--halos sample.}
\label{fig_histo}
\end{figure}

\begin{figure}[]
\hspace{-1.0 cm}
 \resizebox{1.1\hsize}{!}{\includegraphics[clip=true]{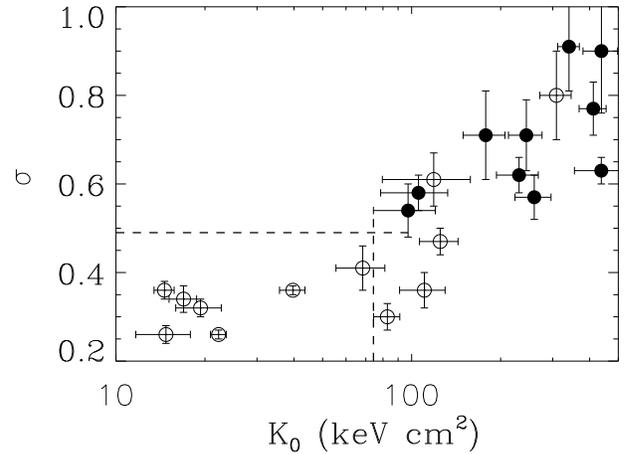}}
\caption{Central entropy and pseudo--entropy ratio for radio--loud (filled circles) and radio--quiet (open circles) clusters. The dashed lines indicate the quadrant of the plot where no-radio halo is found.}
\label{fig_corr}
\end{figure}

In Table \ref{table}, we report $K_0$ and $\sigma$ for all the clusters of our sample.
We classified clusters into core classes according to these indicators. For the sake of simplicity, we decided to classify them into two classes (CC and NCC) well aware of the existence of objects with intermediate properties \citep{leccardi10}. \\
Basing on $K_0$, we divided the clusters population into CC ($K_0<50\,\rm{keV}\,\rm{cm}^2$) and NCC ($K_0>50\,\rm{keV}\,\rm{cm}^2$) as in \citet{cava09}. Using this classification, we found that all ``radio-loud'' clusters are classified as NCC while ``radio quiet'' objects belong to both classes\footnote{A qualitatively similar result was also reported by \citet{ensslin11}.} (Fig.\,\ref{fig_ind} left panel). 
Because of the relatively low number of objects in our sample, we have to verify our result with Monte Carlo simulations to exclude that it comes out just from statistical fluctuations. To this aim, we have calculated the mean $K_0$ of our sample of radio loud clusters ($K_0=274 \pm 14\,\rm{keV}\,\rm{cm}^2$) and compared it with the distribution of the mean $K_0$ of 10 clusters randomly selected in the ACCEPT archive (Fig.\,\ref{fig_histo} left panel). We found that the probability of finding by chance a mean $K_0$ larger than the value of the radio-loud sample is only $0.003\%$ ($0.002\%-0.007\%$ considering the errors on the mean observed value). We have performed the same simulation randomly selecting clusters from the representative HIFLUGS subsample (instead of the whole ACCEPT archive) finding even lower probabilities ($P=10^{-4}\%$), as well as with the subsample of clusters in ACCEPT with redshift in the range $0.2-0.4$ ($P=2\cdot10^{-4}\%$). \\
We have performed the same analysis using the pseudo-entropy ratios $\sigma$, using the threshold  in \citet{rossetti10} to divide objects into classes (CC if $\sigma<0.45$, NCC if $\sigma>0.45$). Again, we found that none of the radio--loud clusters is classified as a CC while radio--quiet objects belong to both classes (Fig.\,\ref{fig_ind} right panel). As for $K_0$, we have performed a Monte Carlo simulation (Fig.\,\ref{fig_histo}), calculating the mean of our ``radio-loud'' sample ($\sigma=0.69 \pm 0.02$) and comparing it with the distribution of the mean of 10 randomly selected values in the sample of \citet{leccardi10}. We found a chance probability of finding a mean value larger than the observed value of $0.26\%$ ($0.02\%-1.96\%$ if we consider the errors on the mean observed $\sigma$).\\
Finally, in Fig.\,\ref{fig_corr} we plot the objects of our sample in the $\sigma-K_0$ plane.  No radio--loud cluster is found in the lower left quadrant of the plot, that we define by subtracting the corresponding error from the minimum $K_0$ ($\sigma$) value of the clusters in the  radio--loud sample (dashed lines in Fig.\,\ref{fig_corr}). We run $10^5$ Monte--Carlo simulations in the $\sigma-K_0$ plane, allowing the points to vary within the error bars and randomly selecting 10 objects between our simulated points.
In only 2 out of $10^5$ simulations, no cluster is found in the lower left quadrant of the plot, proving that our result is statistically significant at $4.3\sigma$. 

\section{Discussion}
\begin{table}
\caption{Statistical results of Monte Carlo simulation.}              
\label{table_MC}      
\centering                                      
\begin{tabular}{c c c}          
\hline\hline          
Test & Value & Null hypothesis probability \\
\hline
Mean $K_0$  & $274\,\rm{keV\,cm^2}$ &  $0.003\%\, (4.17\, \sigma)$ \\
  & $260-288\, \rm{keV\,cm^2}$ &  $0.007-0.002\%\, (3.98-4.26\, \sigma)$ \\
Mean $\sigma$ & $0.69$ & $0.26 \%$, \textbf{($3.01\, \sigma$)}\tablefootmark{\dag} \\
    & $0.67-0.71$ & $1.96-0.02\%\, (2.33-3.71\,\sigma)$ \\
$\sigma$ vs $K_0$ & & $0.002\% (4.26\,\sigma)$\\
\hline
\end{tabular}
\tablefoot{\tablefoottext{\dag} {This is the value we will refer to in the text.}}
\end{table}
As discussed in Sec.\,\ref{sec_results}, we found a robust correlation between the presence of a giant radio halo and the absence of a cool core, as indicated by both $K_0$ and $\sigma$. Despite the relatively low number of objects considered this result is statistically significant as shown by the outputs of our Monte Carlo simulations summarized in Table \ref{table_MC}. Indeed, we found that the probability of a chance result is almost always negligible and lower than $2\%$ even in the worst case. Moreover, we have obtained these results on a well defined sample that unlike archival samples is not biased towards clusters with RHs and without CC (Sec.\,\ref{sec_sample}). \\
One may argue that our choice of excluding the mini-halos clusters from this sample may hamper our results, since this could be exactly the clusters with both a RH and a CC\footnote{We quote here the value of $K_0$ in ACCEPT for the three clusters with mini radio-halos. A2390: $14.73\pm6.99$,
RXCJ$1504.1-0248$: $13.08 \pm 0.94$ and Z7160: $16.88\pm1.52$ \citep{cava09}}. We emphasize here that the classification of these objects as mini-halos was based only on the properties of their radio-emission (e.g. their sizes), regardless of the X-ray properties of their host clusters.
Mini-radio halos are still poorly understood sources (e.g. \citealt{murgia09}), thus we cannot exclude that they could be a phase during the evolution of giant RHs. However, since they are usually considered as a different class of objects from giant RHs with a different origin \citep{mazzotta08, cassano08, zuhone11}, in the following we work under this hypothesis.\\  
The results presented in this paper have important implications on the origin of the CC-NCC dichotomy (Sec. \ref{sec_orig}), since they suggest that the processes leading to the formation of RHs are likely the same that destroy CCs. They also give us the opportunity to estimate the time--scale over which a NCC cluster can relax to the CC state (Sec. \ref{sec_ts}).
\subsection{Origin of the CC-NCC dichotomy.}
\label{sec_orig}
The result presented in this paper are naturally addressed in ``evolutionary'' scenarios of the CC-NCC dichotomy where recent and on-going mergers are responsible for the disruption of the cool cores and for the formation of radio halos. Conversely, alternative ``primordial'' scenarios would have to explain why radio-halos are found only in NCC object.
Radio--halos are transient phenomena with a typical life time of $\sim 1$ Gyr \citep{brunetti09} associated therefore to recent mergers. In the ``primordial'' model of \citet{burns08} early major mergers (at $z>0.5$) destroy nascent cool cores and are responsible for the NCC clusters we observe today. However these early mergers cannot explain the radio emission in most of the radio loud clusters of our sample which have $z\sim 0.2$, corresponding to more than $2.5$ Gyr from $z=0.5$. Even for the most distant cluster of our sample (RXCJ$2003.5-2323$ with $z=0.317$) the life time of the radio halo should be at least $\sim 1.5$ Gyr to reconcile it with the model of \citet{burns08}.
Therefore it is hard to explain the results of the present paper within the ``primordial'' model of \citet{burns08}. It is even harder and against Occam's razor to explain it in the frame-work of ``primordial'' pre-heating models (e.\,g.\, \citealt{mccarthy04}) which require additional physical processes, completely unrelated to those responsible for the radio emission, to account for NCC clusters.

\subsection{Implications on the relaxation time--scale}
\label{sec_ts}
One of the major open issues in the ``evolutionary'' scenario of the CC-NCC dichotomy is the estimate of the likelihood of NCC systems to be transformed into CC objects. The typical cooling times\footnote{We refer to the isobaric cooling time, see discussion in \citet{peterson06}.} in the central regions of NCC are larger than 10 Gyr (e.\,g.\, \citealt{rossetti10}), seeming to imply that once a system has been heated to a NCC state it will not revert to a CC in less than a Hubble time. Indeed, the mean cooling time of the NCC objects ($K_0>50\,\rm{keV}\,\rm{cm^2}$) in our sample is $\simeq 15$ Gyr.\\
We can use the results presented in this paper to estimate the ratio of the two relevant time-scales: the life time of radio-halos ($t_{RH}$) and the time--scale over which NCC relax to the CC state ($t_{NCC\rightarrow CC}$). 
The absence of clusters with RH and CC implies that the two processes leading to the  formation of these objects (i.\,e.\, mergers creating a radio halo but preserving the CC and a fast relaxation from NCC to CC) are extremely unlikely. Indeed the fact that we do not observe this class of clusters but we do observe many NCC objects without a radio halo, implies that RH clusters lose their radio emission more rapidly than developing a new CC, thus $t_{NCC\rightarrow CC}>t_{RH}$. It is possible to show that in stationary conditions this also implies that we cannot have mergers creating a radio halo but preserving the CC (see Appendix A). \\
In this framework, we can provide also an upper limit to the ratio of the two time--scales, which depends on the ratio of the number of radio-quiet clusters with a CC ($ N_{RQ,CC}$)  to those which are NCC  ($ N_{RQ,NCC}$) through the expression:
\begin{equation}
\label{diseq}
1 < \frac{t_{NCC\rightarrow CC}}{t_{RH}}  \leq \frac{N_{RQ,NCC}}{N_{RQ,CC}}\frac{N_{RQ}}{N_{RH}}
\end{equation}
where $N_{RQ}$ and $N_{RH}$ are the total number of radio-quiet clusters and of radio-halos respectively (see Appendix A for the derivation of this expression).
The second fraction in the right term of expression \ref{diseq} is one of the main results of the GMRT RH survey: $N_{RQ}/N_{RH}=25/10$, where we have considered as radio-quiet also the clusters with mini-halos and/or relics, as discussed in Sec. \ref{sec_sample}. Since 10 of the radio-quiet objects in the GMRT RH survey have not been observed by either \xmmn or \chandra, we cannot assess their core state and therefore we cannot really measure the value of the first fraction in the right term of expression \ref{diseq}. If we use $K_0$ as an indicator of the core state the value we measure in our subsample is $N_{RQ,NCC}/N_{RQ,CC}=6/6$ and $6/9$ if we consider also the three mini-halos clusters which are all classified as CC.
Considering the 10 clusters without X-ray observations this ratio could range from $6/19$ to $16/9$. However it is unlikely that the 10 unobserved clusters are either all CC or all NCC. Estimates of the fraction of CC objects in representative samples depend strongly on the indicator used to classify them \citep{chen07,hudson10}. If we consider the HIFLUGS subsample in the ACCEPT catalogue, the fraction of clusters with $K_0<50\,\rm{keV}\,\rm{cm}^2$ is $0.44$. Given that, we expect 6 out of 10 missing clusters to be CC, corresponding to $N_{RQ,NCC}/N_{RQ,CC}=10/15$. Allowing the CC fraction to be in the range $0.35-0.5$ we expect from 3 to 8 CC in the unobserved clusters ($N_{RQ,NCC}/N_{RQ,CC}=8/17$-$13/12$).
Assuming $t_{RH}=1$ Gyr \citep{brunetti09} we find an upper limit $t_{NCC\rightarrow CC}=1.7$ Gyr with the observed ratio $6/9$ and with the expected ratio $10/15$, while in the range $1.2$-$2.7$ Gyr if we allow the ratio to vary between $8/17$ and $13/12$.\\
While we cannot provide more precise constraints because not all clusters in the GMRT RH survey have been observed by \xmmn or \emph{Chandra}, our results show that even in the unlikely case where all the unobserved clusters are NCC, the time scale over which a NCC cluster relaxes to the CC state is less than $4.5$ Gyr.
This time scale is significantly shorter than the typical cooling time of NCC clusters $\simeq 10$ Gyr (e.\,g.\, \citealt{rossetti10}).\\
We can predict that the ratio $N_{RQ,NCC}/N_{RQ,CC}$ should be 4 in the case $t_{NCC\rightarrow CC}=t_{cool} \simeq 10$ Gyr and estimate with a Monte-Carlo simulation that this is is inconsistent at more than $3.4\sigma$ with the permitted values in the GMRT sample assuming a CC fraction in the range $0.35$-$0.5$.\\
As discussed in \citet{rossetti10}, the cooling time derived from thermodynamical quantities should be considered as an upper limit for the relaxation time-scale. 
If, during the merger which destroyed the cool core, the mixing of the gas has not been completely effective and the ICM retains a certain degree of multiphaseness, then the cooler and denser phases will rapidly sink toward the center and re-establish a CC on time--scales shorter than the cooling time calculated under the assumption of uniform temperature and density. Current CCD instruments do not allow us to distinguish a multi-temperature structure with $kT>2$ keV \citep{mazzotta04} and this measurement will become possible only with the high spectral resolution imaging calorimeters on board ASTRO-H. In the meantime, the method we described above provides the opportunity to measure the dynamical time scale $t_{NCC\rightarrow CC}$ and thus to test the possibility of a ``return journey'' to the CC state.

\section{Conclusions}
In this paper, we have analyzed the X-ray observations of the clusters in the GMRT RH survey \citep{venturi07,venturi08}, finding that all clusters which host a radio-halo are also classified as NCC. Although obtained with a relatively low number of objects, this result is statistically significant at more than $3\,\sigma$ (Table \ref{table_MC}).\\
This result implies that the mechanisms which generate radio-halos (most likely mergers) are the same that can destroy cool cores, supporting the ``evolutionary'' origin of the CC-NCC dichotomy. Moreover, we have shown that combining the number of radio--quiet and radio--halos objects with the number of CC and NCC, it is possible to provide upper and lower limits to the ratio of the two relevant time scales: the life time of the radio halo and the relaxation time from NCC to CC.
Assuming $t_{RH}=1$ Gyr \citep{brunetti09}, we constrained $t_{NCC\rightarrow CC}$ in the interval $1-2.7$ Gyr. These values are significantly shorter than the typical cooling time of NCC objects ($t_{cool} \simeq 10$ Gyr, \citealt{rossetti10}), which predicted that most NCC would not develop a new cool-core in less than a Hubble time. On the contrary, the dynamical time--scale we have estimated in this paper allows a ``return journey'' to the CC state and suggests that the gas in the central regions of NCC clusters should be multi-phase. Only with the imaging calorimeters on board ASTRO-H will it be possible to test this important prediction, which may have strong implications on the physics of the ICM.

\begin{acknowledgements}
We would like to thank the referee T.\,Reiprich for his useful and valuable suggestions.
We wish to thank P.\,Humphrey for the use of his \chandra code and K.\,Cavagnolo for the ACCEPT catalogue. 
MR acknowledges stimulating discussions with R.\,Cassano, G.\,Brunetti and T.\,Venturi.
We are supported by ASI-INAF grant I$/009/10/0$, DE is supported also by the Occhialini fellowship at IASF-Milano.
\end{acknowledgements}

\bibliographystyle{aa}

\begin{thebibliography}{42}
\expandafter\ifx\csname natexlab\endcsname\relax\def\natexlab#1{#1}\fi

\bibitem[{{Bacchi} {et~al.}(2003){Bacchi}, {Feretti}, {Giovannini}, \&
  {Govoni}}]{bacchi03}
{Bacchi}, M., {Feretti}, L., {Giovannini}, G., \& {Govoni}, F. 2003, \aap, 400,
  465

\bibitem[{{Brunetti} {et~al.}(2009){Brunetti}, {Cassano}, {Dolag}, \&
  {Setti}}]{brunetti09}
{Brunetti}, G., {Cassano}, R., {Dolag}, K., \& {Setti}, G. 2009, \aap, 507, 661

\bibitem[{{Buote}(2001)}]{buote01}
{Buote}, D.~A. 2001, \apjl, 553, L15

\bibitem[{{Buote}(2002)}]{buote02}
{Buote}, D.~A. 2002, in ASSL Vol. 272: Merging Processes in Galaxy Clusters,
  79--107

\bibitem[{{Burns} {et~al.}(2008){Burns}, {Hallman}, {Gantner}, {Motl}, \&
  {Norman}}]{burns08}
{Burns}, J.~O., {Hallman}, E.~J., {Gantner}, B., {Motl}, P.~M., \& {Norman},
  M.~L. 2008, \apj, 675, 1125

\bibitem[{{Burns} {et~al.}(1997){Burns}, {Loken}, {Gomez}, {Rizza}, {Bliton},
  \& {Ledlow}}]{burns97}
{Burns}, J.~O., {Loken}, C., {Gomez}, P., {et~al.} 1997, in Astronomical
  Society of the Pacific Conference Series, Vol. 115, Galactic Cluster Cooling
  Flows, ed. N.~{Soker}, 21--+

\bibitem[{{Cassano} {et~al.}(2010){Cassano}, {Ettori}, {Giacintucci},
  {Brunetti}, {Markevitch}, {Venturi}, \& {Gitti}}]{cassano10}
{Cassano}, R., {Ettori}, S., {Giacintucci}, S., {et~al.} 2010, \apjl, 721, L82

\bibitem[{{Cassano} {et~al.}(2008){Cassano}, {Gitti}, \&
  {Brunetti}}]{cassano08}
{Cassano}, R., {Gitti}, M., \& {Brunetti}, G. 2008, \aap, 486, L31

\bibitem[{{Cavagnolo} {et~al.}(2009){Cavagnolo}, {Donahue}, {Voit}, \&
  {Sun}}]{cava09}
{Cavagnolo}, K.~W., {Donahue}, M., {Voit}, G.~M., \& {Sun}, M. 2009, \apjs,
  182, 12

\bibitem[{{Chen} {et~al.}(2007){Chen}, {Reiprich}, {B{\"o}hringer}, {Ikebe}, \&
  {Zhang}}]{chen07}
{Chen}, Y., {Reiprich}, T.~H., {B{\"o}hringer}, H., {Ikebe}, Y., \& {Zhang},
  Y.-Y. 2007, \aap, 466, 805

\bibitem[{{Cohn} \& {White}(2005)}]{cohnwhite}
{Cohn}, J.~D. \& {White}, M. 2005, Astroparticle Physics, 24, 316

\bibitem[{{Dickey} \& {Lockman}(1990)}]{dlmap}
{Dickey}, J.~M. \& {Lockman}, F.~J. 1990, \araa, 28, 215

\bibitem[{{Eckert} {et~al.}(2011){Eckert}, {Molendi}, \& {Paltani}}]{eckert11a}
{Eckert}, D., {Molendi}, S., \& {Paltani}, S. 2011, \aap, 526, A79+

\bibitem[{{En{\ss}lin} {et~al.}(2011){En{\ss}lin}, {Pfrommer}, {Miniati}, \&
  {Subramanian}}]{ensslin11}
{En{\ss}lin}, T., {Pfrommer}, C., {Miniati}, F., \& {Subramanian}, K. 2011,
  \aap, 527, A99+

\bibitem[{{Ettori} {et~al.}(2010){Ettori}, {Gastaldello}, {Leccardi},
  {Molendi}, {Rossetti}, {Buote}, \& {Meneghetti}}]{ettori10}
{Ettori}, S., {Gastaldello}, F., {Leccardi}, A., {et~al.} 2010, \aap, 524, A68+

\bibitem[{{Ferrari} {et~al.}(2008){Ferrari}, {Govoni}, {Schindler}, {Bykov}, \&
  {Rephaeli}}]{ferrari08}
{Ferrari}, C., {Govoni}, F., {Schindler}, S., {Bykov}, A.~M., \& {Rephaeli}, Y.
  2008, Space Science Reviews, 134, 93

\bibitem[{{Giacintucci} {et~al.}(2011){Giacintucci}, {Markevitch}, {Brunetti},
  {Cassano}, \& {Venturi}}]{giacintucci11}
{Giacintucci}, S., {Markevitch}, M., {Brunetti}, G., {Cassano}, R., \&
  {Venturi}, T. 2011, \aap, 525, L10+

\bibitem[{{Giovannini} {et~al.}(2009){Giovannini}, {Bonafede}, {Feretti},
  {Govoni}, {Murgia}, {Ferrari}, \& {Monti}}]{giovannini09}
{Giovannini}, G., {Bonafede}, A., {Feretti}, L., {et~al.} 2009, \aap, 507, 1257

\bibitem[{{G{\'o}mez} {et~al.}(2002){G{\'o}mez}, {Loken}, {Roettiger}, \&
  {Burns}}]{gomez02}
{G{\'o}mez}, P.~L., {Loken}, C., {Roettiger}, K., \& {Burns}, J.~O. 2002, \apj,
  569, 122

\bibitem[{{Hudson} {et~al.}(2010){Hudson}, {Mittal}, {Reiprich}, {Nulsen},
  {Andernach}, \& {Sarazin}}]{hudson10}
{Hudson}, D.~S., {Mittal}, R., {Reiprich}, T.~H., {et~al.} 2010, \aap, 513,
  A37+

\bibitem[{{Kalberla} {et~al.}(2005){Kalberla}, {Burton}, {Hartmann}, {Arnal},
  {Bajaja}, {Morras}, \& {P{\"o}ppel}}]{kalberla05}
{Kalberla}, P.~M.~W., {Burton}, W.~B., {Hartmann}, D., {et~al.} 2005, \aap,
  440, 775

\bibitem[{{Kriss} {et~al.}(1983){Kriss}, {Cioffi}, \& {Canizares}}]{kriss83}
{Kriss}, G.~A., {Cioffi}, D.~F., \& {Canizares}, C.~R. 1983, \apj, 272, 439

\bibitem[{{Leccardi} \& {Molendi}(2008)}]{leccardi08a}
{Leccardi}, A. \& {Molendi}, S. 2008, \aap, 486, 359

\bibitem[{{Leccardi} {et~al.}(2010){Leccardi}, {Rossetti}, \&
  {Molendi}}]{leccardi10}
{Leccardi}, A., {Rossetti}, M., \& {Molendi}, S. 2010, \aap, 510, A82+

\bibitem[{{Macario} {et~al.}(2010){Macario}, {Venturi}, {Brunetti},
  {Dallacasa}, {Giacintucci}, {Cassano}, {Bardelli}, \& {Athreya}}]{macario10}
{Macario}, G., {Venturi}, T., {Brunetti}, G., {et~al.} 2010, \aap, 517, A43+

\bibitem[{{Mazzotta} \& {Giacintucci}(2008)}]{mazzotta08}
{Mazzotta}, P. \& {Giacintucci}, S. 2008, \apjl, 675, L9

\bibitem[{{Mazzotta} {et~al.}(2004){Mazzotta}, {Rasia}, {Moscardini}, \&
  {Tormen}}]{mazzotta04}
{Mazzotta}, P., {Rasia}, E., {Moscardini}, L., \& {Tormen}, G. 2004, \mnras,
  354, 10

\bibitem[{{McCarthy} {et~al.}(2008){McCarthy}, {Babul}, {Bower}, \&
  {Balogh}}]{mccarthy08}
{McCarthy}, I.~G., {Babul}, A., {Bower}, R.~G., \& {Balogh}, M.~L. 2008,
  \mnras, 386, 1309

\bibitem[{{McCarthy} {et~al.}(2004){McCarthy}, {Balogh}, {Babul}, {Poole}, \&
  {Horner}}]{mccarthy04}
{McCarthy}, I.~G., {Balogh}, M.~L., {Babul}, A., {Poole}, G.~B., \& {Horner},
  D.~J. 2004, \apj, 613, 811

\bibitem[{{McGlynn} \& {Fabian}(1984)}]{mcglynn84}
{McGlynn}, T.~A. \& {Fabian}, A.~C. 1984, \mnras, 208, 709

\bibitem[{{Murgia} {et~al.}(2009){Murgia}, {Govoni}, {Markevitch}, {Feretti},
  {Giovannini}, {Taylor}, \& {Carretti}}]{murgia09}
{Murgia}, M., {Govoni}, F., {Markevitch}, M., {et~al.} 2009, \aap, 499, 679

\bibitem[{{Peterson} \& {Fabian}(2006)}]{peterson06}
{Peterson}, J.~R. \& {Fabian}, A.~C. 2006, \physrep, 427, 1

\bibitem[{Peterson {et~al.}(2001)Peterson, Paerels, Kaastra, Arnaud, Reiprich,
  Fabian, Mushotzky, Jernigan, \& Sakelliou}]{peterson01}
Peterson, J.~R., Paerels, F. B.~S., Kaastra, J.~S., {et~al.} 2001, \aap, 365, L
  104

\bibitem[{{Rossetti} \& {Molendi}(2010)}]{rossetti10}
{Rossetti}, M. \& {Molendi}, S. 2010, \aap, 510, A83+

\bibitem[{{Sanderson} {et~al.}(2009){Sanderson}, {Edge}, \&
  {Smith}}]{sanderson09b}
{Sanderson}, A.~J.~R., {Edge}, A.~C., \& {Smith}, G.~P. 2009, \mnras, 398, 1698

\bibitem[{{Sanderson} {et~al.}(2006){Sanderson}, {Ponman}, \&
  {O'Sullivan}}]{sanderson06}
{Sanderson}, A.~J.~R., {Ponman}, T.~J., \& {O'Sullivan}, E. 2006, \mnras, 372,
  1496

\bibitem[{{Sarazin}(2002)}]{sarazin02}
{Sarazin}, C.~L. 2002, in ASSL Vol. 272: Merging Processes in Galaxy Clusters,
  1--38

\bibitem[{{Schuecker} {et~al.}(2001){Schuecker}, {B{\"o}hringer}, {Reiprich},
  \& {Feretti}}]{schuecker01}
{Schuecker}, P., {B{\"o}hringer}, H., {Reiprich}, T.~H., \& {Feretti}, L. 2001,
  \aap, 378, 408

\bibitem[{{Snowden} {et~al.}(2008){Snowden}, {Mushotzky}, {Kuntz}, \&
  {Davis}}]{snowden08}
{Snowden}, S.~L., {Mushotzky}, R.~F., {Kuntz}, K.~D., \& {Davis}, D.~S. 2008,
  \aap, 478, 615

\bibitem[{{Venturi} {et~al.}(2007){Venturi}, {Giacintucci}, {Brunetti},
  {Cassano}, {Bardelli}, {Dallacasa}, \& {Setti}}]{venturi07}
{Venturi}, T., {Giacintucci}, S., {Brunetti}, G., {et~al.} 2007, \aap, 463, 937

\bibitem[{{Venturi} {et~al.}(2008){Venturi}, {Giacintucci}, {Dallacasa},
  {Cassano}, {Brunetti}, {Bardelli}, \& {Setti}}]{venturi08}
{Venturi}, T., {Giacintucci}, S., {Dallacasa}, D., {et~al.} 2008, \aap, 484,
  327

\bibitem[{{ZuHone} {et~al.}(2011){ZuHone}, {Markevitch}, \&
  {Brunetti}}]{zuhone11}
{ZuHone}, J., {Markevitch}, M., \& {Brunetti}, G. 2011, in Proceedings of the
  conference "Non-thermal phenomena in colliding galaxy clusters" (Nice, 15-18
  November 2010), astro-ph:1101.4627

\end{thebibliography}

\appendix
\section{Derivation of the time--scale $t_{NCC\rightarrow CC}$}
\begin{figure}[]
 \resizebox{1.1\hsize}{!}{\includegraphics[clip=true, angle=-90]{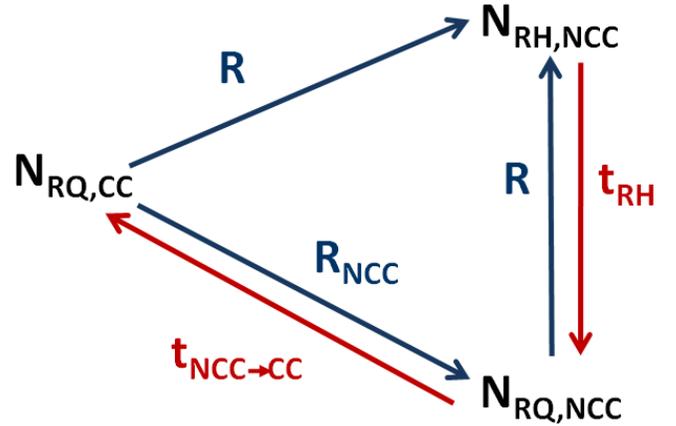}}
\label{fig_3states}
\vspace{-1.0cm}
\caption{A schematic view of the evolution of the thermodynamical and radio state in a cyclical evolutionary scenario. Mergers (blue arrows) are the physical processes that can create a radio-halo or destroy the cool core, while time (red arrows) is responsible for the fading of radio-halos ($t_{RH}$) and for the formation of CCs ($t_{NCC\rightarrow CC}$)}
\end{figure}
The main result of our paper is that there are no RH clusters with a CC.
Therefore we propose a simplified evolutionary scenario (shown in Fig. A.1) where clusters can be found in only three possible states: RH clusters which are also NCC ($N_{RH,NCC}$) and radio--quiet clusters which can be either CC ($N_{RQ,CC}$) or NCC ($N_{RQ,CC}$). We assume that all the mergers that create a RH also destroy the cool core and they happen with a rate $R$. In addition there could be some mergers which destroy the cool core but do not create a RH, happening with a  rate $R_{NCC}$. 
We denote the life-time of the radio-halo as $t_{RH}$ and the time required for a NCC object to relax to the CC state as $t_{NCC\rightarrow CC}$. The underlying assumption (see Sec.\,\ref{sec_ts}) on the time-scales is that $t_{NCC\rightarrow CC} >t_{RH}$, thus clusters with radio halos first lose their radio emission and then develop a new CC.\\
In this scenario, we can write a system of continuity equations for the three states (see \citealt{eckert11a} for a similar argument):
\begin{equation}\left\{\begin{array}{rcl} \frac{dN_{RQ,CC}}{dt} & = & \frac{N_{RQ,NCC}}{t_{NCC\rightarrow CC}}-(R+R_{NCC})N_{RQ,CC} \\
\frac{dN_{RH,NCC}}{dt} & = & R(N_{RQ,CC}+N_{RQ,NCC}) -\frac{N_{RH,NCC}}{t_{RH}}\\
\frac{dN_{RQ,NCC}}{dt} & = & \frac{N_{RH,NCC}}{t_{RH}}+R_{NCC}N_{RQ,CC}-\frac{N_{RQ,NCC}}{t_{NCC\rightarrow CC}}-RN_{RQ,NCC}\end{array}\right.
\label{sys_3}
\end{equation}
Any of these equations is not independent from the other two ($N_{RQ,CC}+N_{RQ,NCC}+N_{RH,NCC}=N_{clusters}$), therefore we decided to keep only the first two equations of the system \ref{sys_3}. We then make the further assumption of a stationary situation, which allows us to cancel the left terms in in expression \ref{sys_3}, and we find:
\begin{equation}\left\{\begin{array}{rcl} \frac{N_{RQ,NCC}}{t_{NCC\rightarrow CC}} & = & (R+R_{NCC})N_{RQ,CC} \\
\frac{N_{RH,NCC}}{t_{RH}} & = & RN_{RQ} \end{array}\right. ,
\label{ex_stat}
\end{equation}
where we used $N_{RQ,NCC}+N_{RQ,CC}=N_{RQ}$. Dividing the first equation by the second, we get
\begin{equation}\frac{N_{RQ,NCC}}{N_{RH,NCC}}=\frac{t_{NCC\rightarrow CC}}{t_{RH}}\left(1+\frac{R_{NCC}}{R}\right)\frac{N_{RQ,CC}}{N_{RQ}}.\end{equation}
Since $R_{NCC}/R\geq 0$,
\begin{equation}\frac{N_{RQ,NCC}}{N_{RH,NCC}}\geq \frac{t_{NCC\rightarrow CC}}{t_{RH}}\frac{N_{RQ,CC}}{N_{RQ}}.
\label{final_ex}
\end{equation}
It is interesting to note that the inequality in the expression \ref{final_ex} becomes an equation when $R_{NCC}=0$, i.\,e.\,if merger events capable of destroying CC always generate radio halos.\\
In principle, we should have considered a four-state system allowing the possibility to have clusters with RH and a CC. We can write the continuity equation for this state, with the usual assumption of a stationary situation:
\begin{equation}
R_{RH,CC}N_{RQ,CC}=R_{NCC}N_{RH,CC}+\frac{N_{RH,CC}}{t_{RH}},
\end{equation}
where $R_{RH,CC}$ is the occurrence of mergers which can produce a RH without destroying the CC.
However, the fact that $N_{RH,CC}=0$, directly implies that $R_{RH,CC}=0$, i.\,e.\, all the mergers capable of generating a RH also destroy the CC, thus justifying the construction of a three-state system such as in Fig.\,A.1\\
One may argue that the key assumption of a stationary situation to derive expression \ref{ex_stat}, is not justified. Indeed the fact that the time scales involved are comparable to
the time interval we are considering should prevent us from assuming equilibrium. However, we recall here 
that in this simplified system we are looking at a ``snapshot'' of an evolutionary process which happens on
time scales of the order of some Gyrs. Unless the merger rate, which is the ultimate factor in determining the changes of state, were to vary abruptly between $z=0.4$ and $z=0.2$, our system should not be very far from equilibrium. Indeed numerical simulations (e.\,g.\, \citealt{cohnwhite}) show that the merger rate should vary only smoothly with time. Moreover, it is possible to take into account the left terms in expression 
\ref{sys_3} and show that if they are smaller (of the order of one tenth) with respect to the change in the number of massive clusters from their formation epoch, our results still hold.
Furtermore, the scenario described here is very simplified also because it assumes that the relevant time scales are the same for all objects while it is more reasonable to expect a distribution of values. If the distributions of $t_{NCC\rightarrow CC}$ and $t_{RH}$ overlap we could have some clusters for which $t_{NCC\rightarrow CC} < t_{RH}$ and therefore we could in principle observe some clusters with both a RH and a CC. Addressing both these issues (the non-stationary situation and the distribution of the time-scales) is not possible with present data, since it requires to follow the evolution of a large sample of clusters over several Gyrs. However, the necessary observations will likely become possible in the next years, thanks to LOFAR and e-ROSITA.

\end{document}